\documentclass{article}
\usepackage[T1]{fontenc} 
\usepackage[utf8]{inputenc} 
\usepackage{ismir,amsmath,cite,url}
\usepackage{graphicx}
\usepackage{color}
\usepackage{subcaption}
\usepackage[ruled,vlined,linesnumbered]{algorithm2e}
\usepackage{booktabs}
\usepackage{multirow}

\usepackage{lineno}

\title{Beat Transformer: Demixed Beat and Downbeat Tracking with Dilated Self-Attention}

\multauthor
{Jingwei Zhao$^{2, 4}$ \hspace{1cm} Gus Xia$^{3, 5}$ \hspace{1cm} Ye Wang$^{1, 2, 4}$} {
$^1$ School of Computing, NUS \hspace{.3cm} $^2$ Institute of Data Science, NUS \hspace{.3cm} $^3$ Music X Lab, NYU Shanghai \\ $^4$ Integrative Sciences and Engineering Programme, NUS Graduate School \hspace{.3cm}$^5$ MBZUAI\\
{\tt\small \href{mailto:jzhao@u.nus.edu}{jzhao@u.nus.edu}, \href{mailto:gxia@nyu.edu}{gxia@nyu.edu}, \href{mailto:wangye@comp.nus.edu.sg}{wangye@comp.nus.edu.sg}}
}

\def\authorname{J. Zhao, G. Xia, and Y. Wang}

\usepackage[bookmarks=false,pdfauthor={\authorname},pdfsubject={\papersubject},hidelinks]{hyperref}

\usepackage{amssymb}
\usepackage{bbm}

\begin{document}

\maketitle
\begin{abstract}
We propose Beat Transformer, a novel Transformer encoder architecture for joint beat and downbeat tracking. Different from previous models that track beats solely based on the spectrogram of an audio mixture, our model deals with \textit{demixed} spectrograms with multiple instrument channels. This is inspired by the fact that humans perceive metrical structures from richer musical contexts, such as chord progression and instrumentation. To this end, we develop a Transformer model with both time-wise attention and instrument-wise attention to capture deep-buried metrical cues. Moreover, our model adopts a novel \textit{dilated self-attention} mechanism, which achieves powerful hierarchical modelling  with only linear complexity. Experiments demonstrate a significant improvement in demixed beat tracking over the non-demixed version. Also, Beat Transformer achieves up to 4\% point improvement in downbeat tracking accuracy over the TCN architectures. We further discover an interpretable attention pattern that mirrors our understanding of hierarchical metrical structures. 
\end{abstract}
\section{Introduction}\label{sec:introduction}

Music audio beat and downbeat tracking, which aims to infer the very basic metrical structure of music, is a long-standing central topic in music information retrieval (MIR). A good beat estimation benefits various downstream MIR tasks, including transcription and structure analysis \cite{holzapfel2016sousta, shibata2020music, nishikimi2020bayesian, ishizuka2021global, donahuesheet}. Moreover, beat tracking can be applied to human-computer interaction \cite{bi2018real, yamamotohuman}, music therapy \cite{DBLP:conf/ismir/CaiEDLW13}, and more scenes, as beats echo with human perceptual and motor sensitivity to musical rhythms.

We see significant progress in beat tracking with the development of deep neural networks. Current mainstream methods utilize temporal convolutional networks (TCNs)  to extract frame-wise beat activations from an input spectrogram \cite{matthewdavies2019temporal}. We further see successful efforts in boosting beat tracking performance, including phase-informed post-processing with dynamic Bayesian networks (DBNs) \cite{krebs2015efficient}, multi-task learning for joint beat, downbeat and tempo estimation \cite{bock2016joint, bock2019multi, bock2020deconstruct}, and explicit beat phase modelling \cite{oyamaphase}. 



Recently, Transformer has demonstrated highly competitive performances over a range of MIR tasks \cite{hung2022modeling, lu2021spectnt, hawthorne2021sequence, ou2022exploring, gardner2021mt3, won2019toward, won2021semi}. In this paper, we propose Beat Transformer, a novel Transformer encoder architecture for joint beat and downbeat tracking. To better accommodate Transformer to our purpose, we introduce two extra inductive biases. Firstly, our model is constructed with \textit{short-windowed dilated self-attention}. An exponentially increasing dilation rate enables our model to discern beats from non-beats in a hierarchical manner. With a fixed window size, our model maintains a linear complexity to the input sequence length.



Another inductive bias is \textit{demixed beat tracking}. This strategy is inspired by the fact that human beat tracking is always accompanied by and enhanced by a deep understanding of the musical contexts. For example, the coordination of instruments enforces the progression of chords and bass notes, thus implying metrical accents, and such cues can be easily identified by human listeners. To capture this relation, we use Spleeter \cite{spleeter2020} to demix an input music piece into multiple instrument channels, and our model performs both \textit{time-wise} and \textit{instrument-wise} attention in alternate Transformer layers to excavate metrical cues.

We evaluate Beat Transformer on a wide range of beat- and downbeat-annotated datasets. Besides competing with state-of-the-art works, we present a thorough ablation study to illustrate the effectiveness of dilated self-attention and demixing. Moreover, our model learns highly interpretable representations. We demonstrate that our model can be interpreted as a learner over finite-state Markov chains, and we observe beat phase transition through visualization of the transition (attention) matrix. 

In brief, the contributions of our paper are as follows:
\begin{itemize}
    \item We propose Beat Transformer\footnote{Available at \href{https://github.com/zhaojw1998/Beat-Transformer}{https://github.com/zhaojw1998/Beat-Transformer}.}, a novel Transformer encoder architecture for joint beat and downbeat tracking in music audio.
    \item We devise dilated self-attention, which demonstrates powerful sequential modelling with linear complexity, potentially adaptable to more general MIR tasks.
    \item We make use of music demixing to complement and enhance beat tracking, shedding light on future MIR research towards universal music understanding.
\end{itemize}

\section{Related Works}
We review two topics related to our work: beat tracking, and Transformer. For beat tracking, we focus on its development with deep neural networks. For Transformer, we address its application in MIR. For more general review of both topics, we refer readers to \cite{tempobeatdownbeat} and \cite{lin2021survey}, respectively.

\subsection{Music Audio Beat Tracking}
Beat tracking has been formulated as a two-stage sequential learning task. The first stage aims to determine the likelihood of beat presence, or beat activation, at each frame of an input spectrogram. The initial deep learning approach for this purpose was based on long short-term memory networks (LSTM) \cite{bock2011enhanced, bock2014multi}. To better pick up the beat sequence from raw model output, at the second stage, a dynamic Baysian network (DBN) is introduced to infer tempo and beat phase transition from beat activation \cite{krebs2015efficient}. 

The current mainstream methods substitute the LSTM with a TCN architecture \cite{matthewdavies2019temporal, bock2019multi, bock2020deconstruct, oyamaphase, hung2022modeling}. Specifically, the convolutional kernels have a dilation rate exponential to the depth of layer. This hierarchical structure facilitates the network to model various scales, functionally similar to pooling, but maintains the same input and output size \cite{oord2016wavenet}. Besides architecture, another breakthrough of beat tracking is the formulation of multi-task learning \cite{bock2016joint, bock2019multi, bock2020deconstruct, oyamaphase, hung2022modeling}. Specifically, beat, downbeat, and tempo are strongly correlated metrical features. Sharing model weights among all three sub-tasks helps each to reach better convergence.

A recent trend of beat tracking is to deal with \textit{demixed} music sources. Chiu \textit{et. al.} leverages demixed drum and non-drum streams to enhance model adaptability to different drum source conditions \cite{chiu2021source, chiu2021drum}. In fact, humans can track beats while switching their attention among different instrument parts. Hence the coordination of instrumental sources can be explored for useful metrical information. 

In our work, we inherit the fashion of multi-task learning and the use of DBN, while proposing a novel Transformer encoder architecture to replace TCN. We formalize \textit{dilated self-attention} \cite{dai2021pdan, moritz2021capturing} for efficient modeling of long metrical structures.  Moreover, we seek to enhance beat tracking by introducing \textit{instrumental attention} among drum, piano, bass, vocal, and other demixed sources. 


\subsection{Transformer in MIR}
Transformer has established itself \textit{de facto} state-of-the-art in natural language and symbolic music domain. Recently, it has also demonstrated outstanding performance over audio-based MIR tasks, including music transcription \cite{hawthorne2021sequence, ou2022exploring, gardner2021mt3}, music tagging \cite{won2019toward, won2021semi}, and other analysis \cite{lu2021spectnt}. For these tasks, a Transformer model is trained over short spectrogram clips typically of 2-5 seconds as a compromise to the quadratic complexity computing self-attention.

Transformer is first applied to beat tracking by Hung \textit{et. al.} \cite{hung2022modeling} using SpecTNT blocks \cite{lu2021spectnt}. For each SpecTNT block, a spectral Transformer encoder first aggregates spectral information at each time step, and then a temporal encoder exchanges information in time and pays attention to beat and downbeat positions. Such a time-frequency design makes an effective use of spectral features and achieves the state-of-the-art performance.

Our work is also a Transformer-based architecture that strives to enhance beat tracking with richer musical contexts. Instead of aggregating spectral features as in \cite{hung2022modeling}, we resort to demixed instrumental attention as a more explicit inductive bias to exploring spectral information. Our design of dilated self-attention is also a crucial step to accommodate Transformer to beat tracking. With only linear complexity, our model handles full-length songs at a time.



\begin{figure*}[htb]
     \centering
     \begin{subfigure}[b]{.7\textwidth}
         \centering
         \includegraphics[width=\textwidth]{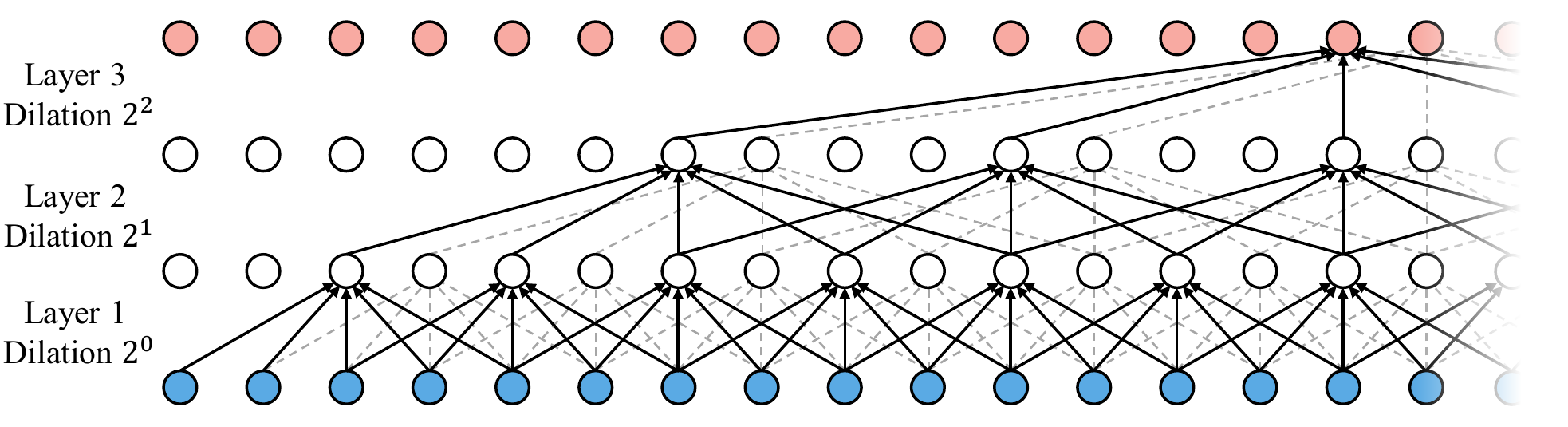}
         \caption{Layer-wise view of dilated self-attention (Adapted from TCN structures \cite{matthewdavies2019temporal, oord2016wavenet})}
         \label{dsa_layer_view}
     \end{subfigure}
     \begin{subfigure}[b]{.28\textwidth}
         \centering
         \includegraphics[width=\textwidth]{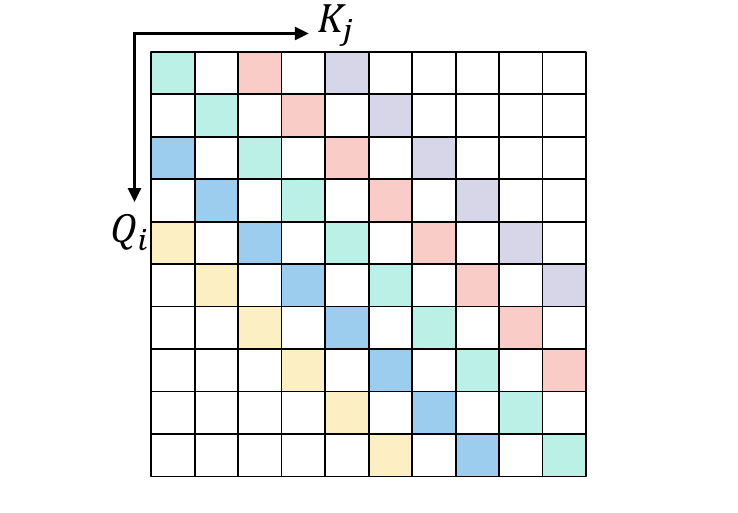}
         \caption{Attention matrix view at layer 2}
         \label{dsa_matrix_view}
     \end{subfigure}
     \caption{Illustration of dilated self-attention (with a non-causal short window of size 5) over a three-layer Transformer. Part (a) shows the hierarchical connectivity across layers, which shares the same pattern as TCN in \cite{matthewdavies2019temporal}. Part (b) shows the attention matrix at layer 2, with colours indicating relative position. The white colour indicates unattainable positions.}
     \label{illus_dsa}
\end{figure*}

\section{Method}
The core of our method is a Transformer encoder based on \textit{1) dilated self-attention} (DSA), and \textit{2) demixed instrumental attention}, to extract a framewise beat activation from input spectrograms. In this section, we first formalize DSA in Section \ref{dsa}. We then present our design of demixed beat tracking in Section \ref{demixed}. We further interpret our method with Markov chain properties in Section \ref{markov}.

\subsection{Dilated Self-Attention (DSA)}\label{dsa}
\subsubsection{Background of Self-Attention (SA)}
We first recall that, for vanilla Transformer layers, self-attention (SA) is computed via the scaled dot product:
\begin{equation}\label{vanila_sa}
    \mathrm{Attention}(Q, K, V) = \mathrm{softmax}(\frac{QK^{\top}}{\sqrt{d_\mathrm{f}}})V,
\end{equation}
where $Q_{1:T}$, $K_{1:T}$, $V_{1:T} \in \mathbb{R}^{T\times d_\mathrm{f}}$ are query, key, and value sequences, each linearly mapped from input $x_{1: T}$. $T$ is the sequence length, and $d_\mathrm{f}$ is the feature dimension.

The partial attention from position $i$ to $j$ is explicitly:
\begin{equation}\label{qk_sa}
    e_{ij} = \frac{Q_i K_j^\top}{\sqrt{d_\mathrm{f}}},
\end{equation}
where  $1 \leq i, j \leq T$. Such computation leads to quadratic complexity $\mathcal{O}(T^2)$ in terms of both time and space.

\subsubsection{Dilated Self-Attention (DSA)}\label{dsa_idea}

An illustration of DSA is shown in Figure \ref{illus_dsa}. DSA is computed over a short window of size $l_\mathrm{{win}}=m+n+1$, where $m$ and $n$ are the length of non-causal and causal components of the window (in Figure \ref{illus_dsa}, $m=n=2$). Each Transformer layer has a dilation rate $r \geqslant 1$, and $r$ increases exponentially as the layer goes deeper. 

Formally, given $Q$, $K$, and $V \in \mathbb{R}^{T\times d_\mathrm{f}}$, DSA first computes Q-K attention by:
\begin{equation}\label{dsa1}
    e_{ik} = \frac{Q_i K_{i + rk}^\top}{\sqrt{d_\mathrm{f}}},
\end{equation}
where $1\leq i \leq T$ and $-m \leq k \leq n$. Specifically, $i+rk$ refers to the positions in $K_{1: T}$ that are attainable by $Q_i$ under the dilated window of rate $r$. When $i+rk$ exceeds the sequence range $[1, T]$, we fill $e_{ik}$ with $\mathtt{-inf}$.

Then, the Q-K attention $e_{ik}$ is normalized via softmax:
\begin{equation}\label{dsa2}
    p_{ik} = \frac{\exp(e_{ik})}{\sum_{k=-m}^n \exp(e_{ik})}
\end{equation}

The output of DSA is a sequence $z_{1: T}$, where $z_i$ is the weighted average of $V$ under the same dilated window:
\begin{equation}\label{dsa3}
    z_i = \sum_{k=-m}^n p_{ik}(V_{i + rk})
\end{equation}

We apply DSA with exponentially increasing dilation rates and relative positional embedding (RPE) \cite{shaw2018self} to a stack of Transformer layers. Each layer consists of two sub-layers: DSA, and a position-wise feed-forward layer. We place residual connections across each sub-layer and perform layer normalization \cite{ba2016layer} before each sub-layer.






\subsubsection{Memory Efficient Implementation of DSA}\label{implement-dsa}
\begin{figure}
 \centerline{
 \includegraphics[width=0.9\columnwidth]{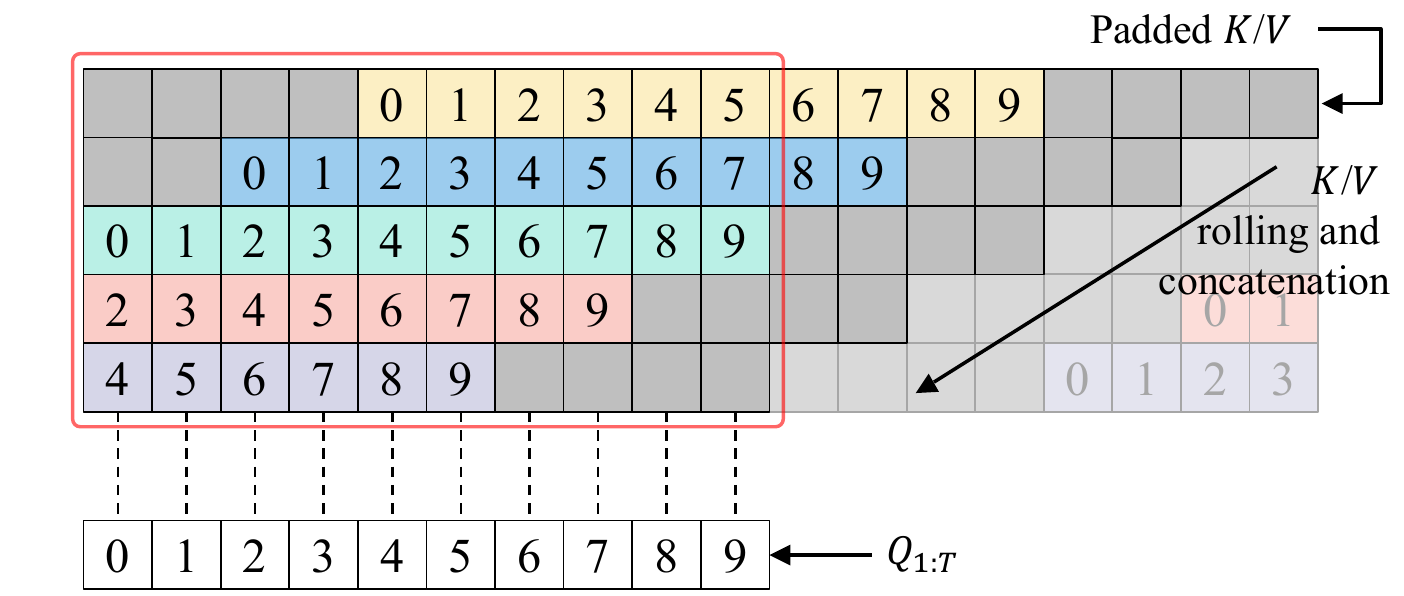}}
 \caption{Illustration of efficient DSA implementation, with dilation rate $r=2$ and window size $l_\mathrm{win}=5$.}
 \label{efficient_dsa}
\end{figure}

A straightforward implementation of DSA is to mask SA. Concretely, a square mask takes the same form as in Figure \ref{dsa_matrix_view}, where all uncolored positions are filled with $\mathtt{-inf}$, rendering a rather sparse attention matrix. This way, however, still requires quadratic complexity, because $e_{ij}$ is explicitly computed for all masked positions.

Our implementation takes a ``rolling'' strategy to eliminate redundant computation. As in Figure \ref{efficient_dsa}, $l_\mathrm{win}$ copies of $K_{1:T}$ sequences are padded and rolled along the time axis, and then concatenated on a new axis. In this way, each $Q_i$ sees $K_j$ directly and only at $j=i+rk$ for $-m\leq k \leq n$, which is exactly the coverage of the dilated window. 

Formally, given $K_{1: T} \in \mathbb{R}^{T\times d_\mathrm{f}}$, dilation rate $r$, and window size $l_\mathrm{win}$, the rolling strategy takes three steps:
\begin{enumerate}
    \item Pad $K_{1:T}$ with $\lfloor \frac{l_\mathrm{win}}{2} \rfloor \times r$ steps on both sides;
    \item Make $l_\mathrm{win}$ copies of padded $K$. Starting from 0, each copy is cyclically rolled $r$ more steps;
    \item Concatenate each copy along a new axis and retrieve the first $T$ steps. The output has shape $T\times l_\mathrm{win} \times d_\mathrm{f}$.
\end{enumerate}

The same procedure applies to $V_{1:T}$ as well. In this way, computing DSA is essentially as simple as Equation \eqref{vanila_sa}. Here, instead of $Q, K, V \in \mathbb{R}^{T\times d_\mathrm{f}}$, we have $Q \in \mathbb{R}^{T\times 1 \times d_\mathrm{f}}$ and $K$, $V \in \mathbb{R}^{T\times l_\mathrm{win} \times d_\mathrm{f}}$, with $T$ treated as a batch dimension. The computation complexity is $\mathcal{O}(T\times l_\mathrm{win})$, while $l_\mathrm{win}$ is fixed and small enough to be left uncounted.

    

\subsection{Demixed Beat Tracking}\label{demixed}
\begin{figure}
 \centerline{
 \includegraphics[width=0.9\columnwidth]{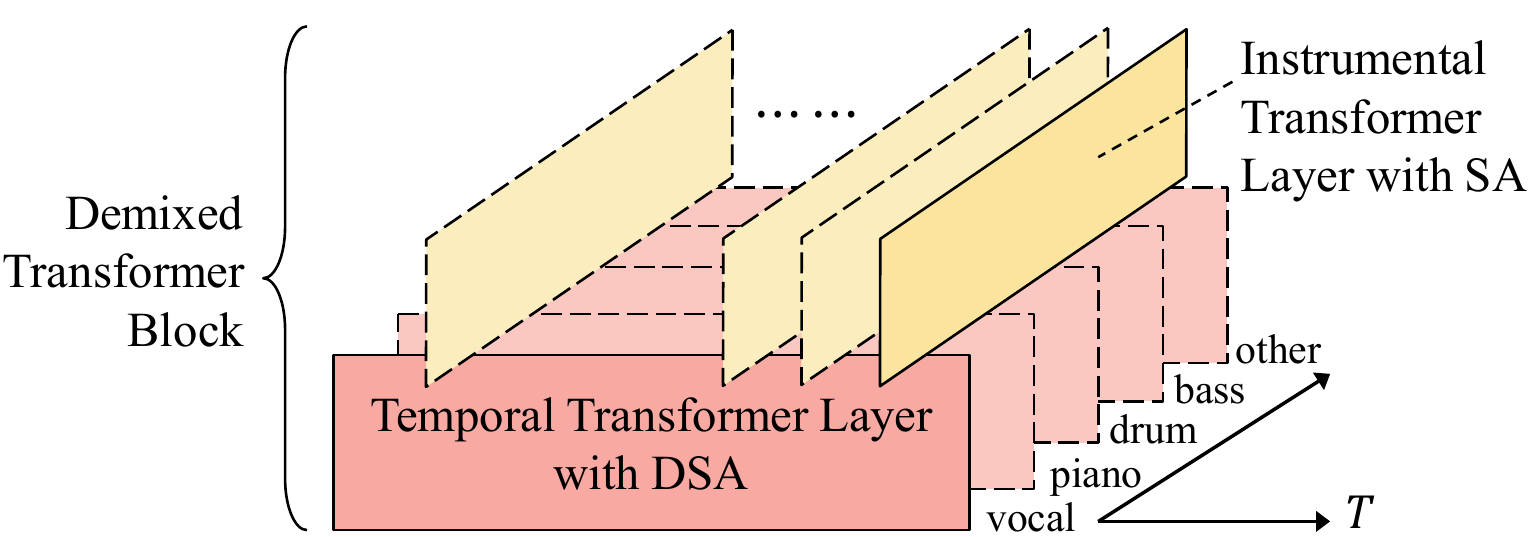}}
 \caption{Demixed Transformer block. Two Transformer layers are stacked ``orthogonally'', each handling time-wise dilated self-attention and instrument-wise self-attention.}
 \label{instr_attn}
\end{figure}

\begin{figure*}
 \centerline{
 \includegraphics[width=1.9\columnwidth]{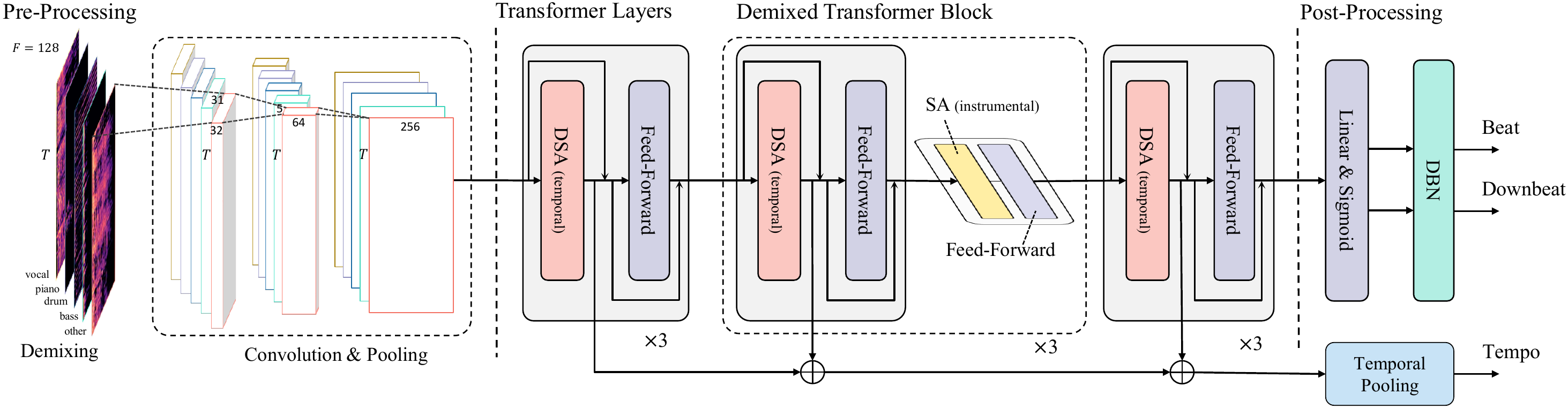}}
 \caption{Beat Transformer architecture. For conciseness, layer normalization and dropout layers are not shown.}
 \label{full_model}
\end{figure*}

\subsubsection{Demixed Transformer Block}
We use Spleeter 5-stems model \cite{spleeter2020} to demix an input piece into spectrograms with $\left| C \right|$ instrument channels, where $C = \{ \mathtt{vocal}, \mathtt{piano}, \mathtt{drum}, \mathtt{bass}, \mathtt{other} \}$. As shown in Figure \ref{instr_attn}, we stack two Transformer layers to perform time-wise and instrument-wise attention in turn. Let the input at layer $l$ be $x^{l}_{1: T, 1: \left| C \right|} \in \mathbb{R}^{T\times \left| C \right| \times d_\mathrm{f}}$, a \textit{temporal} Transformer layer (TTL) first takes $x_{1: T, c}^{l}$ for $1 \leq c \leq  \left| C \right|$:
\begin{equation}
    x_{1: T, c}^{l+1} = \mathrm{TTL}(x_{1: T, c}^{l})
\end{equation}

 Then, an \textit{instrumental} Transformer layer (ITL), on the \textit{orthogonal} direction, takes $x_{t, 1: \left| C \right|}^{l+1}$ for $1 \leq t \leq T$:
\begin{equation}
    x_{t, 1: \left| C \right|}^{l+2} = \mathrm{ITL}(x_{t, 1: \left| C \right|}^{l+1})
\end{equation}

A TTL followed by ITL forms a \textit{demixed Transformer block}. TTL consists of DSA as described in Section \ref{dsa_idea}. For ITL, we use vanilla SA because there is only 5 instrument channels. As instruments are not sequentially ordered, we do not add any positional encoding to ITL. 

Through demixed Transformer blocks, our model can capture the rhythmic evolution of each instrument, as well as the harmonic coordination among all instruments. 

\subsubsection{Partial Demix Augmentation}\label{augmentation}

Spleeter may produce empty channels when a certain instrument does not present. To avoid potential effects of such a situations, we develop a partial demix strategy for data augmentation. Partial demix creates new stems by summing up existing instrument channels of a default 5-stem demixed input sample. For example, an augmented data sample may have three channels corresponding to $C^{\prime} = \{ \mathtt{vocal\&piano}, \mathtt{drum}, \mathtt{bass\&other} \}$.  

In our case, we randomly sum up $2$, $3$, or $4$ instrument channels of a 5-stem input with a probability $30\%$, $10\%$, and $10\%$ during training. In this way, our model is encouraged to pay attention to \textit{instrument-agnostic} musical contents and thus is less affected by empty channels where no valid music content is present. As our augmentation strategy also adds to the demix diversity and the data quantity, we believe it brings general benefits to training as well.

 
\subsection{Markov Chain Interpretation}\label{markov}
In Equation \eqref{dsa2}, we formulate the attention matrix of DSA as $P=[p_{ij}]_{1 \leq i, j \leq T}$, where $p_{ij} \geq 0$ if and only if $j = i + rk$ for $-m \leq k \leq n$. Here, $r$ is the dilation rate, and $m$, $n$ are components of the attention window. Moreover, $P$ satisfies $\sum_{j=1}^T p_{ij} = 1$ for all $i$. Therefore, $P$ can be regarded as the transition matrix of a finite-state Markov chain, where each state is a position of the input sequence.

For a stack of temporal Transformer layers (TTL), where DSA is employed, layer $l$ essentially learns a unique one-step transition $P^{l}$, by which our model can attend to local neighbours covered by the attention window. Through $L$ layers, our model makes an $L$-step transition, during which it attends to global positions hierarchically. The overall $L$-step transition matrix $P^{(L)}$ satisfies:
\begin{equation}\label{product_markov}
P^{(L)} = \prod_{l=1}^{L}P^{l}
\end{equation}

Note that $P^{l}$ itself is a rather sparse matrix (due to \textit{short} attention window), while ${P^{(L)}}$ is densely connected. Its components $[p^{(L)}_{ij}]_{1 \leq i, j \leq T}$ represent the hierarchical attention weights across the whole $L$ layers, which can tell us much richer attention patterns (more in Section \ref{attn_visual}).

\begin{table*}[h]
  \centering
  \fontsize{5}{5}\selectfont
  \resizebox{\textwidth}{!}{
  
    \begin{tabular}{cccccccc}
    \toprule
          &       & \multicolumn{3}{c}{\textbf{Beat Accuracy}} & \multicolumn{3}{c}{\textbf{Downbeat Accuracy}} \\
\cmidrule{3-8}    \textbf{Dataset} & \textbf{Model} & \textbf{F-Measure} & \textbf{CMLt} & \textbf{AMLt} & \textbf{F-Measure} & \textbf{CMLt} & \textbf{AMLt} \\
    \midrule
    \multirow{6}[2]{*}{Ballroom} & TCN+Demix & 0.960 & 0.942 & 0.960 & 0.925 & 0.924 & 0.956 \\
          & Ours w/o Demix & 0.968 & 0.946 & 0.965 & 0.930 & 0.925 & 0.963 \\
          & Ours w/o Aug. & 0.967 & 0.949 & \underline{\textbf{0.967}} & 0.928 & 0.931 & 0.958 \\
          & Ours  & \underline{\textbf{0.968}} & \underline{\textbf{0.954}} & 0.966 & \underline{\textbf{0.941}} & \underline{\textbf{0.944}} & \underline{\textbf{0.969}} \\
          & B{\"o}ck \textit{et al.} \cite{bock2020deconstruct} & 0.962 & 0.947 & 0.961 & 0.916 & 0.913 & 0.960 \\
          & Hung \textit{et al.} \cite{hung2022modeling} & 0.962 & 0.939 & 0.967 & 0.937 & 0.927 & 0.968 \\
    \midrule
    \multirow{6}[2]{*}{Hainsworth} & TCN+Demix & 0.887 & 0.827 & 0.918 & 0.739 & 0.708 & \underline{0.861} \\
          & Ours w/o Demix & 0.902 & \underline{0.844} & \underline{0.934} & 0.721 & 0.688 & 0.843 \\
          & Ours w/o Aug. & 0.892 & 0.831 & 0.908 & 0.742 & 0.703 & 0.837 \\
          & Ours  & \underline{0.902} & 0.842 & 0.918 & \underline{\textbf{0.748}} & \underline{0.712} & 0.841 \\
          & B{\"o}ck \textit{et al.} \cite{bock2020deconstruct} & \textbf{0.904} & 0.851 & \textbf{0.937} & 0.722 & 0.696 & \textbf{0.872} \\
          & Hung \textit{et al.} \cite{hung2022modeling} & 0.877 & \textbf{0.862} & 0.915 & 0.748 & \textbf{0.738} & 0.870 \\
    \midrule
    \multirow{6}[2]{*}{Harmonix} & TCN+Demix & 0.954 & 0.903 & 0.956 & \underline{0.901} & \underline{0.866} & \underline{0.923} \\
          & Ours w/o Demix & 0.954 & 0.902 & \underline{0.958} & 0.887 & 0.846 & 0.916 \\
          & Ours w/o Aug. & 0.952 & 0.901 & 0.950 & 0.897 & 0.863 & 0.919 \\
          & Ours  & \underline{\textbf{0.954}} & \underline{0.905} & 0.957 & 0.898 & 0.863 & 0.919 \\
          & B{\"o}ck \textit{et al.} \cite{bock2020deconstruct}$^*$ & 0.933 & 0.841 & 0.938 & 0.804 & 0.747 & 0.873 \\
          & Hung \textit{et al.} \cite{hung2022modeling} & 0.953 & \textbf{0.939} & \textbf{0.959} & \textbf{0.908} & \textbf{0.872} & \textbf{0.928} \\
    \midrule
    \multirow{6}[2]{*}{SMC} & TCN+Demix & 0.596 & 0.455 & 0.625 &       &       &  \\
          & Ours w/o Demix & 0.589 & 0.448 & 0.621 &       &       &  \\
          & Ours w/o Aug. & 0.595 & 0.450 & 0.626 &       &       &  \\
          & Ours  & \underline{0.596} & \underline{0.456} & \underline{0.635} &       &       &  \\
          & B{\"o}ck \textit{et al.} \cite{bock2020deconstruct} & 0.552 & 0.465 & 0.643 &       &       &  \\
          & Hung \textit{et al.} \cite{hung2022modeling} & \textbf{0.605} & \textbf{0.514} & \textbf{0.663} &       &       &  \\
    \midrule
    \multirow{6}[2]{*}{GTZAN} & TCN+Demix & 0.873 & 0.780 & 0.907 & 0.700 & 0.646 & 0.842 \\
          & Ours w/o Demix & 0.876 & 0.787 & 0.914 & 0.686 & 0.633 & 0.834 \\
          & Ours w/o Aug. & 0.881 & 0.797 & 0.921 & 0.703 & 0.653 & \underline{0.845} \\
          & Ours  & \underline{0.885} & \underline{0.800} & \underline{0.922} & \underline{0.714} & \underline{0.665} & 0.844 \\
          & B{\"o}ck \textit{et al.} \cite{bock2020deconstruct} & 0.885 & \textbf{0.813} & \textbf{0.931} & 0.672 & 0.640 & 0.832 \\
          & Hung \textit{et al.} \cite{hung2022modeling} & \textbf{0.887} & 0.812 & 0.920 & \textbf{0.756} & \textbf{0.715} & \textbf{0.881} \\
    \bottomrule
    \end{tabular}%
    }
    \caption{Testing results of beat and downbeat tracking under 8-fold cross-validation. GTZAN is unseen from training and held out for test only. B{\"o}ck \textit{et. al.} \cite{bock2020deconstruct} on Harmonix is reproduced by \cite{hung2022modeling}, as indicated by the $^*$ symbol. We use \underline{underscore} to denote best results comparing with our ablation models and use \textbf{boldface} to compare with state-of-the-art models.}
  \label{testing_results}%
\end{table*}%

\subsection{Complete Architecture}
A complete view of Beat Transformer is presented in Figure \ref{full_model}. The inputs are log-scaled spectrograms demixed by Spleeter, with $F=128$ mel-bins and $\left| C \right|=5$ instrument channels. Subject to Spleeter, our frame rate is $43.07$ fps and the frequency range is up to 11 kHz. We then use three 2D convolutional layers, shared by each demixed channel, as a front-end feature extractor. The convolutional design is the same as in \cite{bock2020deconstruct} except that we employ more filters to reach feature dimension $d_\mathrm{model}=256$.

Beat Transformer comprises 9 temporal Transformer layers (TTL) with DSA. Each TTL has 8 attention heads with window size $l_\mathrm{win}=5$, four of which have skewed window ranges, where $m=0, 1, 3, 4$, respectively, and $n=4-m$. The dilation rate grows exponentially from $2^0$ to $2^8$, stretching to a receptive field of $47.51$ seconds. Among the 9 TTLs, the middle three are expanded to demixed Transformer blocks by interleaving ITLs. We found it sufficient to perform instrumental attention only in the middle layers, which has a proper scale of $1$-$5$ seconds. Each Transformer layer has 8 heads ($d_\mathrm{f}=32$) followed by a feed-forward layer with hidden dimension $d_\mathrm{ff}=1024$. 

We sum up the instrument channels of the output of the last Transformer layer and obtain a frame-wise beat representation of shape $T\times d_\mathrm{model}$. Following multi-task learning practice \cite{bock2016joint, bock2019multi, bock2020deconstruct, oyamaphase}, we use a linear layer to map the representation to beat and downbeat activations respectively, and add a regularization branch predicting global tempo via ``skip connections'' \cite{bock2019multi}. We apply DBN in Madmom package \cite{bock2016madmom} as the post-processor to pick up the beat and downbeat sequence from raw activations. For DBN parameters, we set $\mathtt{observation\_lambda}=6$, $\mathtt{transition\_lambda}=100$, and $\mathtt{threshold}=0.2$.

\section{Experiments}
\subsection{Datasets}
We utilize a total of 7 datasets for model training and evaluation: \textit{Ballroom} \cite{gouyon2006experimental, krebs2013rhythmic}, \textit{Hainsworth} \cite{hainsworth2004particle}, \textit{RWC Popular} \cite{goto2002rwc}, \textit{Harmonix} \cite{nieto2019harmonix}, \textit{Carnetic} \cite{srinivasamurthy2014supervised}, \textit{SMC} \cite{holzapfel2012selective}, and \textit{GTZAN} \cite{tzanetakis2002musical, marchand2015gtzan}. We acquire \textit{Harmonix} in mel-spectrogram and invert each piece to audio using Griffin-Lim Algorithm \cite{griffin1984signal, perraudin2013fast} with Librosa package \cite{mcfee2015librosa}. Following convention, we leave \textit{GTZAN} for testing only and use the other datasets in 8-fold cross validation \cite{bock2016joint, bock2019multi, bock2020deconstruct}.

\subsection{Training}
Our model is supervised in a multi-task learning fashion, where beat, downbeat, and tempo are predicted jointly \cite{bock2020deconstruct}. Beat and downbeat annotations are each represented as a 1D binary sequence that indicates beat ($1$) and non-beat ($0$) states at each input frame.  Following \cite{bock2020deconstruct}, we widen beat and downbeat states to $\pm2$ neighbours of annotated frames with weights $0.5$ and $0.25$. Following \cite{bock2019multi}, we derive tempo target from beat annotation for the tempo prediction branch. We found the use of tempo branch generally beneficial to beat tracking, as it may serve as a regularization term that helps reaching better convergence.

For training, we combine the binary cross entropy loss over beat, downbeat, and tempo by weighing them equally. We use a batch size of 1 to train on whole sequences with different lengths. For excessively long songs, we split them into 3-minute (8k-frame) clips. We apply RAdam \cite{liu2019variance} plus Lookahead \cite{zhang2019lookahead} optimizer with an initial learning rate of $1\mathrm{e}{-3}$, which is reduced by a factor of 5 whenever the validation loss gets stuck for 2 epochs before being capped at a minimum value of $1\mathrm{e}{-7}$. We use dropout \cite{srivastava2014dropout} with rate $0.5$ for the tempo branch and $0.1$ for other parts of the network. We apply \textit{partial demix augmentation} described in Section \ref{augmentation} as the only means of data augmentation. Our model has 9.29M trainable parameters and is trained with an RTX-A5000-24GB GPU. Each training fold generally takes 20 epochs (in 11 hours) to fully converge. 

\begin{figure*}
     \centering
     \begin{subfigure}[b]{.21\textwidth}
         \centering
         \includegraphics[width=\textwidth]{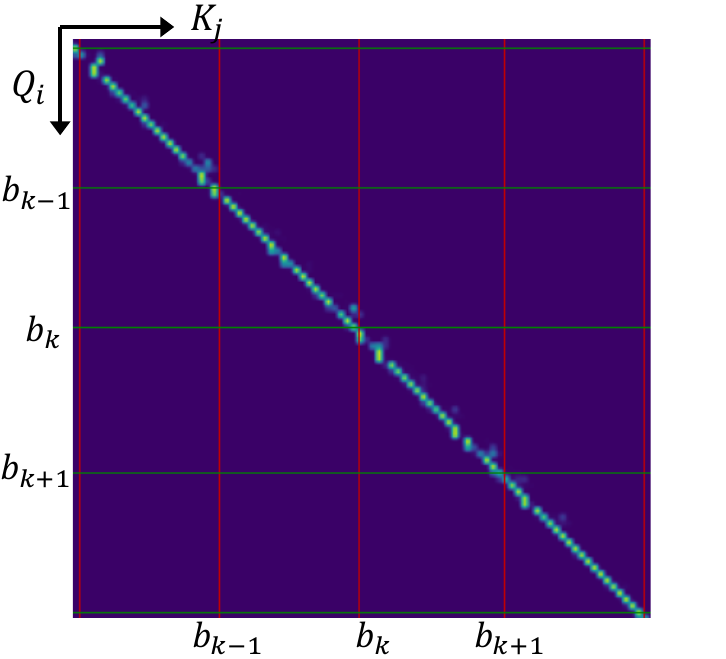}
         \caption{${P^{(1)}}$ lower scale}
         \label{visual1}
     \end{subfigure}
     \begin{subfigure}[b]{.21\textwidth}
         \centering
         \includegraphics[width=\textwidth]{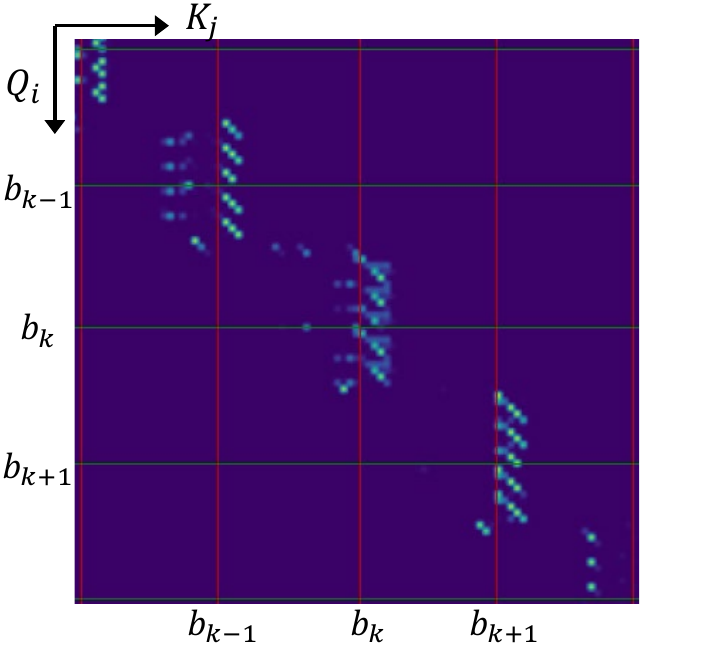}
         \caption{${P^{(3)}}$ beat scale}
         \label{visual2}
     \end{subfigure}
     \begin{subfigure}[b]{.21\textwidth}
         \centering
         \includegraphics[width=\textwidth]{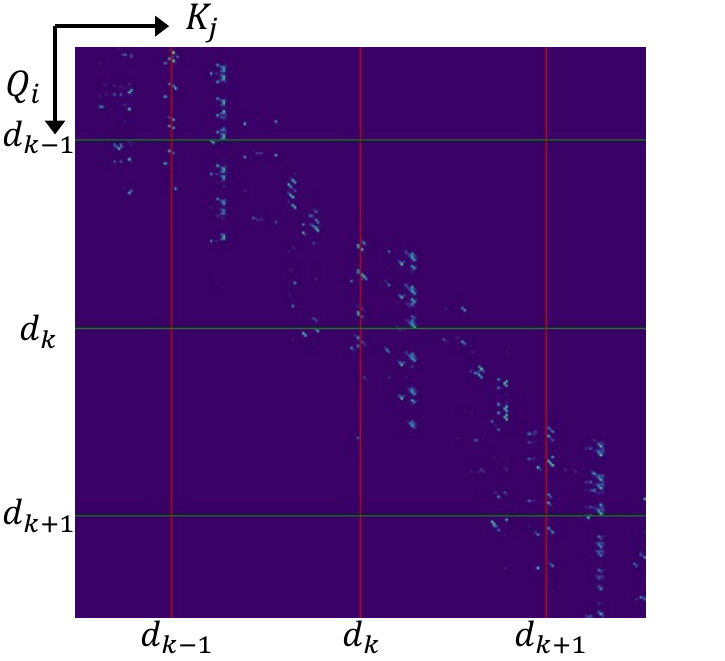}
         \caption{${P^{(5)}}$ downbeat scale}
         \label{visual3}
     \end{subfigure}
     \begin{subfigure}[b]{.21\textwidth}
         \centering
         \includegraphics[width=\textwidth]{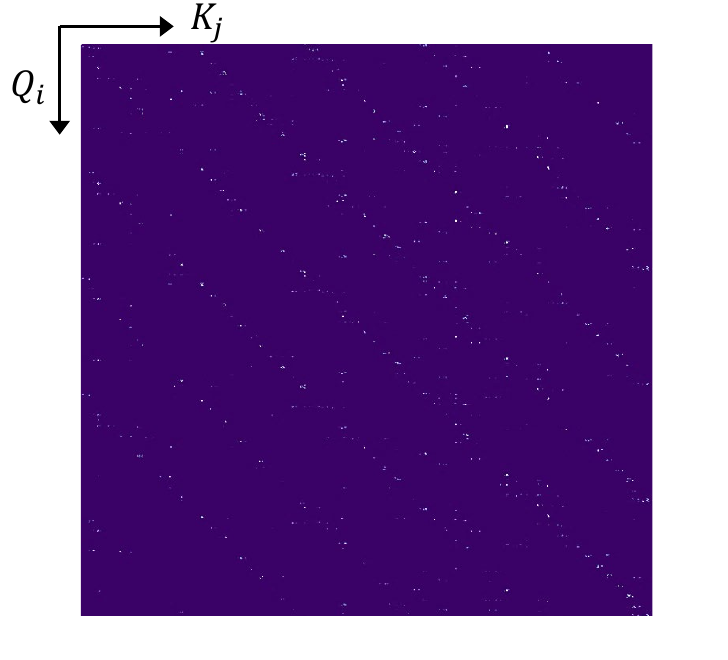}
         \caption{${P^{(9)}}$ higher scale}
         \label{visual4}
     \end{subfigure}
     \caption{Visualization of temporal attention matrix based on the product rule in Equation \eqref{product_markov} for $L$-step transition on a Markov chain. $L=1, 3, 5, $ and $9$ from part (a) to (d), which model different hierarchies of the metrical structure.}
     \label{visual}
\end{figure*}

\subsection{Evaluation}
\subsubsection{Baseline Methods}

We first compare three ablation models to validate our module design. The first ablation model is \textit{Ours} trained without partial demix augmentation (\textit{\textbf{Ours w/o Aug.}}). The second model removes all ITL layers and tracks beat with a single channel of non-demixed mixture (\textit{\textbf{Ours w/o Demix}}). The last model replaces each TTL layer with a TCN layer \cite{bock2020deconstruct} of the same dilation rate (\textit{\textbf{TCN+Demix}}). We implement TCN layers following \cite{tempobeatdownbeat} while setting the input and output shape to be the same as our model. The resulting model yields a comparable amount of 10.25M parameters and is trained without augmentation.

In addition, we compare our model with two recent works that have achieved state-of-the-art performance. Specifically, B{\"o}ck \textit{et al.} \cite{bock2020deconstruct} is based on TCN architectures and Hung \textit{et al.} \cite{hung2022modeling} is based on SpecTNT.


\subsubsection{Results and Discussion}

In Table \ref{testing_results}, we first observe \textit{Ours w/o Aug.} yields generally better performance than \textit{TCN+Demix}, especially on the unseen \textit{GTZAN} dataset. As both models share a comparable amount of parameters, this result demonstrates the capability of Transformer (DSA) versus TCN (dilated convolution), which also corroborates with previous findings on Transformer's comparability to convolution on general tasks \cite{dosovitskiy2020image, touvron2021training, liu2021swin}. Considering that Transformer is notoriously data-inefficient to train, it is remarkable that our model is well-trained with limited data without augmentation. We owe this merit to DSA, which not only prevents redundant computation but also makes musical sense in terms of the hierarchical structure of music metrical modelling.

Comparing \textit{Ours w/o Demix} to \textit{Ours w/o Aug.}, while both models are highly competitive in beat tracking, the latter demonstrates more superiority in downbeat tracking. Downbeat tracking is generally more difficult than beat tracking because it is involved with deeper musical knowledge, such as chord and bass progression, behind the apparent spectrogram energy. In our model, the instrumental attention captures the instrumental coordination as hints to the harmonic cues that are orthogonal to the temporal axis, and thus acquires better metrical modelling. 

Comparing \textit{Ours w/o Aug.} to \textit{Ours}, we observe a consistent improvement across datasets brought by partial demix augmentation, which indicates the general usefulness of this augmentation strategy to model training.

Compared to state-of-the-art models, our improvement in downbeat accuracy is more significant than that in beat accuracy. On the test-only \textit{GTZAN} dataset, we obtain 4\% point gain in \textit{F-measure} over B{\"o}ck \textit{et al.} \cite{bock2020deconstruct} in downbeat tracking. Compared to Hung \textit{et al.} \cite{hung2022modeling}, which is also based on Transformer, our model can be more flexibly trained (owing to the efficient DSA mechanism) on a 24GB GPU in contrast to four 32GB GPUs reported in \cite{hung2022modeling}.

\subsection{Attention Matrix Visualization}\label{attn_visual}
We visualize the attention matrix that our model learns by interpreting it as a \textit{multi-step Markov transition matrix} as defined in Equation \eqref{product_markov}. Specifically, the $L$-step matrix is the product of $L$ one-step matrices through $L$ layers. Here we only consider TTLs with dilated self-attention, as ITLs work on an orthogonal axis. Figure \ref{visual} shows the attention matrix $P^{(L)}$ for $L=1, 3, 5, $ and $9$ of the $\mathtt{drum}$ channel, inferred from the piece $\mathtt{hiphop.00090}$ chosen from \textit{GTZAN}.

Figure \ref{visual1} shows a one-step transition. Each position $Q_i$ can only attend to its neighbours covered by the attention window. Still, we observe that beat positions (denoted by $b_{k}$) are likely to get more attention. In Figure \ref{visual2} where we step to the beat scale, most attention spots are aligned with beats. Moreover, we observe that the attention at $b_{k}$ is typically prolonged after $Q_i$ leaves $b_{k}$, and is formed before $Q_i$ reaches $b_{k}$ for every $k$. This means that our model learns to transition its attention from the \textit{offbeat} phase following the last beat to the \textit{upbeat phase} preceding the next beat. Figure \ref{visual3} further stretches the view to the downbeat scale, and we can see similar patterns aligned with downbeat positions (denoted by $d_{k}$). Finally, in Figure \ref{visual4}, the attention reaches further positions and displays a structural pattern.

The above visualization demonstrates the inner logic that our model exploits for beat and downbeat tracking. We see that our model gathers information from both local and global scales with an organized hierarchy. 

\section{Conclusion}
In conclusion, we contribute a novel Transformer architecture for audio beat and downbeat tracking. The main novelty lies first in our design of dilated self-attention, which brings down the computation complexity of Transformer from quadratic to linear level. In addition, we successfully enhance beat and downbeat tracking by utilizing off-the-shelf progress in music demixing. Our model not only captures deeper harmonic cues for better metrical inference but also discerns beat and downbeat in a visualizable hierarchical manner. Our model is efficient, interpretable, and potentially generalizable with highly competitive sequential modelling power. We hope our model encourages future MIR research toward universal music understanding.

\bibliography{ISMIRtemplate}

\end{document}